# Ultracompact on-chip silicon optical logic gates

Chen Wang and Zhi-Yuan Li[1]

**All-optical integrated circuits for computing and information processing have been pursued for decades as a potential strategy to overcome the speed limitations intrinsic to electronics. However feasible on-chip integrated logic units and devices still have been limited by its size, quality, scalability, and reliability. Here we demonstrate all-passive on-chip optical AND and NAND logic gates made from a directional emitting cavity connecting two ultrasmall photonic crystal heterojunction diodes. The measured transmission spectra show more than 10dB contrast of the logic transport with a high phase tolerance, agreeing well with numerical simulations. The building of linear, passive, and ultracompact silicon optical logic gates might pave the way to construct novel nanophotonic on-chip processor architectures for future optical computing technologies.**

The semiconductor-based electronics is rapidly approaching its fundamental limit caused by interconnect delays and large heat generation. Yet, the pace of seeking the next-generation communication technologies never slows down. Light-based devices have been proposed as a potential strategy for advancing semiconductor-based computing beyond the fundamental performance limitations of electronic devices, as epitomized by Moore's law[1,2]. Much as logic devices play the basic and critical part in electronics, all-optical logic devices are an indispensible part of modern integrated optical circuits and chips. Previous developments have focused on two routes to construct optical logic gates: the first one is based on linear optical effects, such as interferometry[3], semiconductor optical amplifier (SOA) and Mach–Zehnder interferometer (MZI)[4,5]. This type of optical logic devices is based on mature techniques of optical fibers but it requires large space and is impossible for micro- or nano-integration. In micro-structures linear effect all-optical logic functions, such as optical interference effect[6], were demonstrated, which preliminarily solves the integration problem. But the disadvantage of this type of optical logic devices is the lack of polarization and phase tolerance, indicating it requires precise initial conditions that are very hard to implement. The interference plasmonic networks[7] were utilized to implement optical logic functions but the plasmonic loss problem limits its further application in cascaded logic gates. The second route is based on nonlinear processes. Tunable refractive index or frequency mixing were proposed[8,9], but they are subject to the size problem and not suitable for integration. Switching effects can be used to realize nonlinear all-optical logic gates in nano-structures[10], where input pump signals play as switches along the light path under the nonlinear effect and control the status of output signals for logic functions. This type of logic devices includes electro-optical effect[11,12], thermal-optical effect[13], two-photon absorption[14] and the third-order nonlinear effect[15,16]. These devices have high performance but its high power needs and slow response time cannot meet the requirement of large-scale optical computing.

---

[1] Laboratory of Optical Physics, Institute of Physics, Chinese Academy of Sciences, P. O. Box 603, Beijing 100190, China

A more stringent issue of optical logic devices is their compatibility with conventional complementary metal oxide semiconductor (CMOS) processing[18-20] for the sake of practical applications. Silicon photonics has been widely assumed as a good platform for optical integration. Previous functional silicon optical logic devices are also based on linear optical effects[6] or nonlinear processes[14-16], especially in photonic crystal (PC) structures[20]. However, there exist intrinsic difficulties of modulating light using silicon structures because of the weak dependence of the refractive index and absorption coefficient on the free-carrier concentration in silicon. To date, all optical logical devices built on silicon have relatively large lengths (on the order of millimetres). In addition, they also strongly request high power or external electromagnetic field. The ultimate goal to find a kind of low power and easy-to-integrate optical logic devices in silicon is still far away. To realize this fundamental purpose, new concepts must be explored. How about going back to electronic system to find the inspiration?

As is known, electronic logic devices are basically the combination of various electronic diodes[21]. For example, an AND gate is usually designed using two cascaded diodes [Fig. 1(a)]. The digital inputs A and B cause the output L to have the result as the AND function. If neither or only one input to the AND gate is HIGH, a LOW output results. Other logic functions can also be achieved by different assemblies of electronic diodes. So a natural question is asked: Are there any possibilities to combine optical diodes to build optical logic gates?

Recently an ultrasmall on-chip optical diode based on silicon photonic crystal heterojunction was reported[22,23]. The optical diode is linear, passive, and time-independent, but has a spatial-inversion symmetry breaking geometry. The overall size of the ultrasmall diode is only 6×6 μm$^2$. Here we demonstrate all-passive ultracompact on-chip logic gates by assembling two of the passive optical diodes [Fig. 1(b)]. The logic operations are realized by the directional emitting cavity buried in the wavelength-scale heterojunction without nonlinearity or magnetism. A particularly important step is the construction of the AND and NAND gates, and more complex logic functions and functional photonic circuits can be fulfilled by cascading the basic gates.

Figure 2(a) shows the schematic configuration of the AND logic gate structure under study, which includes two diodes and a connecting cavity. Each diode consists of two photonic crystal slab domains (PC1 and PC2) with the same lattice constant $a$ but different air hole radii ($r_1$ and $r_2$, respectively) comprising a heterojunction structure. These two photonic crystal regions stand at a silicon slab [grey area in Fig. 2(a)]. Each photonic crystal region has a square-lattice pattern of air holes [white holes in Fig. 2(a)], with the hetero-interface between PC1 and PC2 along the Γ-M direction. The hole radius of PC1 and PC2 are $r_1 = 0.24a$ and $r_2 = 0.36a$. The center cavity joins the two diodes together and provides a directional emission to the diodes, which eventually plays a basic role as the logic functions. The light source is placed symmetrically aside the left of the structure with two 2$a$-wide ridge waveguides, as the input ports, connecting the center cavity. The logic signal is collected at the right of the structure with two 4$a$-wide ridge waveguide, as the output port, connecting one of the diode waveguide regions. The AND logic function is defined as the intensity of the output signal in dependence on the status of the intensity of the two input signals. For the AND gate structure, it is expected that the output signal has a much higher intensity at the presence of both input signals than those when only one input signal in present.

To see the performance of the designed logic gate structure, we first simulated the transmission spectra for a TE-like light signal, which includes $E_x$, $E_y$ and $H_z$ electromagnetic components, transporting along the *x*-axis direction using the three-dimensional finite-difference time-domain (3D-FDTD) method[24]. Figure 2(b) shows the calculated output transmission spectra signal with only one input port used (red line for only input port A and blue line for only input port B) and both two input ports used (black line). It is clearly seen that there exists a logic band ranging from 0.266 to 0.285 ($a/\lambda$), where the output signals with only one input port used are merely -26 dB while the two-port-active output signal reaches -16 dB. This indicates the two-port signal is one order of magnitude larger than the single-port signal, which has already caught up with the standard electronic logic devices. We define $S = T_{AB} - \max(T_A, T_B)$ as the signal contrast of the logic device, where $T_{AB}$, $T_A$ and $T_B$ denote the two-port-active and single-port output transmissivity, respectively. The maximum S of this AND logic device equals 10.5 dB at 0.2741 ($a/\lambda$).

Based on the above numerical analysis, the AND logic gate structure was fabricated in a silicon slab by nanofabrication techniques. Figure 2(c) shows the scanning electron microscopy image of the fabricated sample along the light path. The lattice constant *a* was 440 nm, the radii $r_1$ and $r_2$ of air holes in the PC1 and PC2 were approximately 110 nm and 160 nm, respectively. The slab thickness was 220 nm (h = 0.5*a*). To measure the transmission spectra, infrared light from a semiconductor laser, which is tunable between 1,500 nm and 1,640 nm, was directly coupled into the photonic crystal region with the aid of tapered ridge waveguides in the input and output ends[25]. The input power is set to 5 mW with -10 dB direct coupling loss between fibers and ridge waveguides. The wavelength region is normalized to 0.2683-0.2933 (*a*/λ), enough to encompass the AND logic band.

Figures 2(d) shows the experimental results of the transmission spectra of the AND logic device. The experimental transmissions are optimized as the input/output loss has been removed against a reference sample. In Fig. 2(d) the experimental two-port-active output signal at 1,605 nm has the maximum transmissivity of -15.3 dB, whereas the single-port-active output signal has a transmissivity of -28.9 dB and -30.7 dB, respectively, in good agreement with the theoretical maximum transmissivity in Fig. 2(b). The measured signal contrast S equals 13.6 dB. The experiment confirms the existence of the AND logic effect in agreement with the theoretical prediction. Due to the arbitrariness of the lattice constant *a*, the logic frequency can be freely adjusted to anywhere as desired. This could be very convenient for the design of realistic photonic devices. Another prominent point is the low power level of the logic structure. In experiment the laser power used in the ridge waveguides and the main structure is merely 0.284 MW/cm$^2$ [ ≈ 0.5mW/(800nm×220nm)], which totally cannot afford any remarkable nonlinear effects of the silicon material[26], indicating that the logic structure is completely linear and passive.

To have a deeper understanding of the extraordinary AND logic function in this passive and linear silicon structure, the detailed mode distribution of the logic gate structure is studied. The structure can be divided into two parts: the center cavity region [pink area in Fig. 3(a)] and the output photonic crystal diode waveguide region [black area in Fig. 3(a)]. The role of the center cavity is to bring a connection between the two diodes and implement a logic function, while the role of the output photonic crystal diode waveguide region is to convert the logic signal to the propagation mode of the waveguide and output it to the ridge waveguide. To confirm this assumption, we calculate the resonance spectrum of the cavity by using the 3D FDTD method in association with the Pade approximation[24] for the TE-like modes. The spectrum is obtained by taking a discrete Fourier transform of a 42000 time-step ($3.75 \times 10^{-17}$ s for every time step) sequence initialized with a broad-band source. We found that there is a rarely-low-Q (around 20) cavity mode, whose peak is located at 0.2818 ($a/\lambda$) and covers the whole logic frequencies. Field distribution with point source [Fig. 3(b)] shows that this cavity has very strong leakage, which shows an "X"-shape field profile, corresponding with its low Q factor. The output photonic crystal waveguide region [black line in Fig. 3(a)] is collecting the up-right branch of the leaky mode. The field distributions using the real input signals are also calculated [Figs. 3(c)-(e)]. When single input is used [Figs. 3(c), (d)], only two of the leaky branches are active [left two branches in Fig. 3(c) and down two branches in Fig. 3(d)], but none are active in the up-right branch that could output because the input signal only has parallel or vertical **k** vector of light. When two inputs are both active, the **k** vector in the cavity has a strong component along the up-right branch [Fig. 3(e)] so that the output is strongly enhanced. Note that the mixture of the **k** vectors happens after the input signals are injected and stored in the cavity reserve, as a result of which the phase condition of the input signal should be irrelevant to the logic signal.

Because the input signal must go through the center emitting cavity, it is expected that the passive logic device has natural advantages of phase sensitivity tolerance. To confirm this assumption, we simulate the two-port-active output signal with different phase delays. First we use the Gaussian pulse waves, the same as used in Fig. 2(b), to create different phase delays. The result [Fig. 4(a)] shows that even if the two light sources have an initial phase difference of 0.5π, 1.0π and 1.2π, the output signals are the same as the one with no phase difference. Then we change the light sources into continuous waves at 0.2818 ($a/\lambda$) with the initial phase difference of 0.5π, 1.0π and 1.2π, and the result [Fig. 4(b)] remains the same as that with no phase difference. These calculation results clearly indicate that the coupling of the center cavity could erase the phase information of the input signal and enhance the tolerance of phase sensitivity. We further do experimental test and fabricate a sample with an arbitrary phase delay [Fig. 4(d)] to compare with the no-phase-delay structure [Fig. 4(c)]. The length difference ΔL between input port A and input port B is 349 nm so that the phase difference $\Delta\varphi(=\frac{2\pi}{\lambda} \bullet n_{eff} \bullet \Delta L)$ ranges from 1.1π to 1.2π for different wavelengths. The experimental result in Fig. 4(e) fully supports our simulation results and shows that the logic structure has remarkable tolerance of phase sensitivity.

We further change the order of the light propagation, from PC1 waveguide to the next directional emitting cavity, so that a NAND logic device is prepared. Figures 5(a) and 5(b) show the schematic geometry and the scanning electron microscopy images of the fabricated NAND logic structures along the light path. The lattice constant $a$ was 370 nm, the radii $r_1$ and $r_2$ of air holes in the two photonic crystals still follow $r_1 = 0.24a$ and $r_2 = 0.36a$, respectively. The slab thickness $h$ was $0.5a$. Unlike the AND logic gate, the working frequency from the diode waveguide is different from the AND logic gate, which is the reason why the lattice constant $a$ is changed to fit our laser in experiment, so that the emitting cavity, which also shows an "X"-shape field profile, now chooses different leaky branches that decrease the output signal when two ports are both active. The measured transmission spectra in Fig. 5(d), showing apparent NAND logic transport of light, agree well with numerical simulations in Fig. 5(c). In experiment, we have got the maximum transmissivity of the two-port-active output signal -38.2 dB, whereas the single-port-active output signals are of a transmissivity of -16.4 dB and -20.5 dB, and a signal contrast S of -17.7 dB at 1,578 nm.

The principle for optical logic device as analyzed in the above is robust as it is only based on a simple directional emission effect of a cavity embedded within the photonic crystal heterojunction. The cavity makes the device insensitive to phase variation of the two input signals and only depend on their intensities. As a result, the optical logic devices function in the same robust way as the electronic logic devices do. Similar principles can be adopted to realize other basic logic function gates.

However, there exist problems of remarkable loss in the designed optical logic devices. The loss can consist of the insertion loss between ridge waveguides (input ports) and the photonic crystal region and the connection loss between the cavity with input and output ports. The former one is found to be about 8.9 dB for AND gate and the same for NAND gate. Optimization can be made to reduce remarkably these values. In these situations, the absolute output signal intensity could reach 22.9% (-6.4 dB) for the AND gate and 17.8% (-7.5 dB) for the NAND gate of the input signal intensity. The intrinsic loss could be solved by further optimizing the geometries of logic gates and by adding other active modules in the photonic network.

The all-optical silicon AND and NAND logic gates work on the directional emitting cavity connecting two photonic crystal heterojunction diodes. They rely neither on nonlinearity nor on magnetism, have high signal contrast and wavelength-scale ultrasmall sizes, and are all-dielectric, linear, and passive. More importantly, they have high polarization and phase tolerances, and are compatible with CMOS processing technologies. All these unique features can greatly facilitate large-scale integration. More complicated logic networks can be built based on these basic logic gates and other photonic crystal structures[25], which may pave the way to construct tomorrow's on-chip large-scale logic circuits and optical computers.

**Methods**

Simulation: We use 3D-FDTD method for numerical simulation on the transmission spectra and field patterns. The doubled-diode structure is used to calculate the round-trip transmission spectrum and the single diode structure is used to calculate the forward and backward transmission spectra, which all contain the same input/output ridge waveguide. The width of the input waveguide is the $2a$ ($a$ =440 nm is the lattice constant) and the output waveguide is $4a$. The grid size used in the simulation is 11 nm. The light source is placed at the input port with $2a$-wide ridge waveguides connecting the surface of the diode region and the receiving plane is placed at the output port. The whole area is surrounded by a perfectly matched layer.

Sample fabrication: The patterns were first defined in resist using the electron beam lithography (EBL) on the top layer of a silicon-on-insulator (SOI) chip. The resist patterns were then transferred to silicon layer using the inductive coupled plasma reactive ion etching (ICP-RIE) technique. The insulator layer ($SiO_2$) underneath the silicon pattern regions was finally removed by a HF solution to form an air-bridged structure.

Measurements: In our optical experiments, light from a semiconductor laser, which is tunable between 1,500 nm and 1,640 nm, was directly coupled from a single mode tapered fiber into the input ridge waveguides at the input port. And at the output port, a single mode tapered fiber was used to attach the output ridge waveguides and the power meter by direct coupling. We also measured the transmission power of the reference waveguide, which had only the same tapered ridge waveguides but no diode region. The loss from the input fiber to the output fiber is -20 dB.

**Acknowledgements**

This work was supported by the National Basic Research Foundation of China under grant no. 2011CB922002 and Knowledge Innovation Program of the Chinese Academy of Sciences (No. Y1V2013L11).

**Author Information** Reprints and permissions information is available at www.nature.com/reprints. Correspondence and requests for materials should be addressed to Z.Y. L. (lizy@aphy.iphy.ac.cn).

**Author Contributions**

Z.Y.L. designed the study, C.W. performed the numerical simulations, fabricated the samples, collected, and analyzed the data. Z.Y.L. and C.W. wrote the paper. All authors discussed the results and commented on the manuscript.

**Figures:**

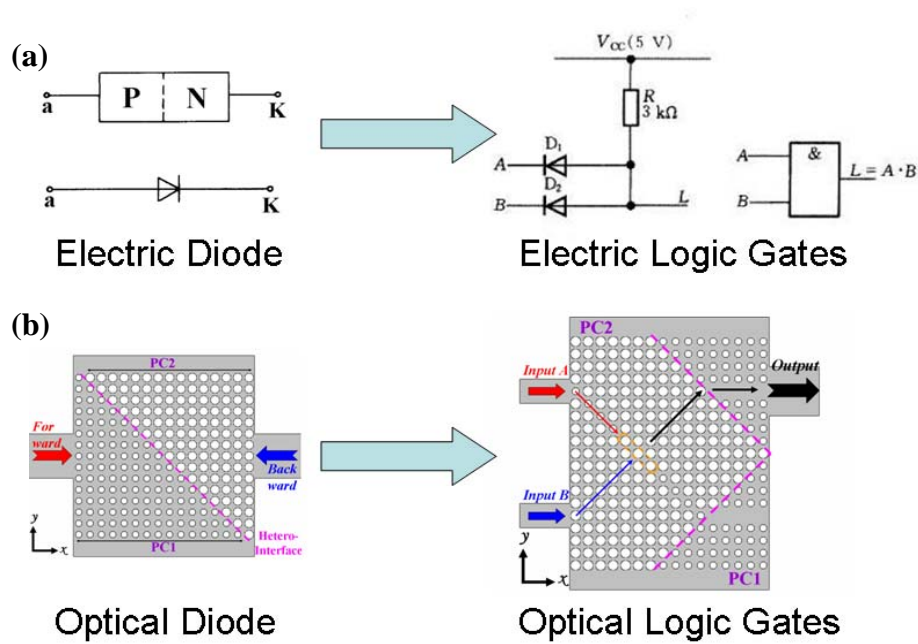

**Figure 1 | Construction of the electronic and optical logic gates.**
Schematic geometry of the developments (a) from electronic diodes to electronic logic gates, and (b) from optical diodes to optical logic gates. The optical logic devices could be realized by the combination of optical diodes, in much similar way as the electronic logic devices based on the combination of various electronic diodes.

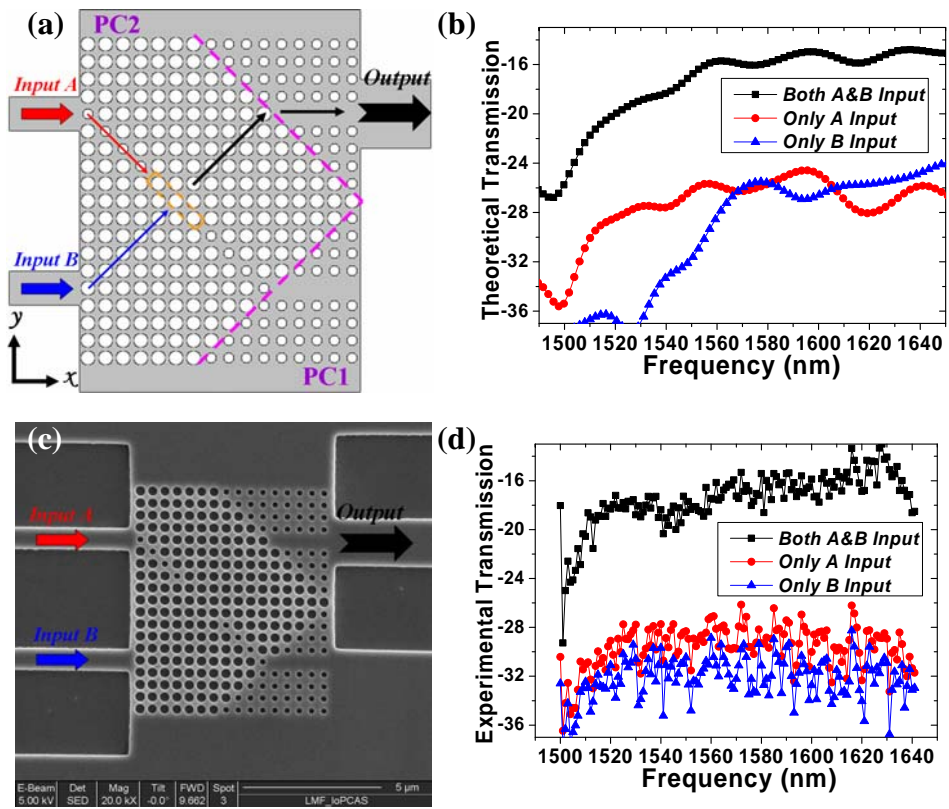

**Figure 2 | Demonstration of the optical AND logic gate.**
(a) Schematic geometry of the AND logic gate formed by a cavity connecting two diodes with interfaces (normal to the Γ-M direction) between two photonic crystal slabs (denoted as PC1 and PC2) with different hole radii ($r_1$ and $r_2$, respectively). (b) Simulated transmission spectra of the AND logic gate at the output port with single input (the red and blue lines) and with both two inputs (the black line). (c) Scanning electron microscope images, and (d) experimental transmission spectra of AND logic gate structure. The measured transmission spectra, showing more than 10dB contrast of the logic transport of light, agree well with numerical simulations.

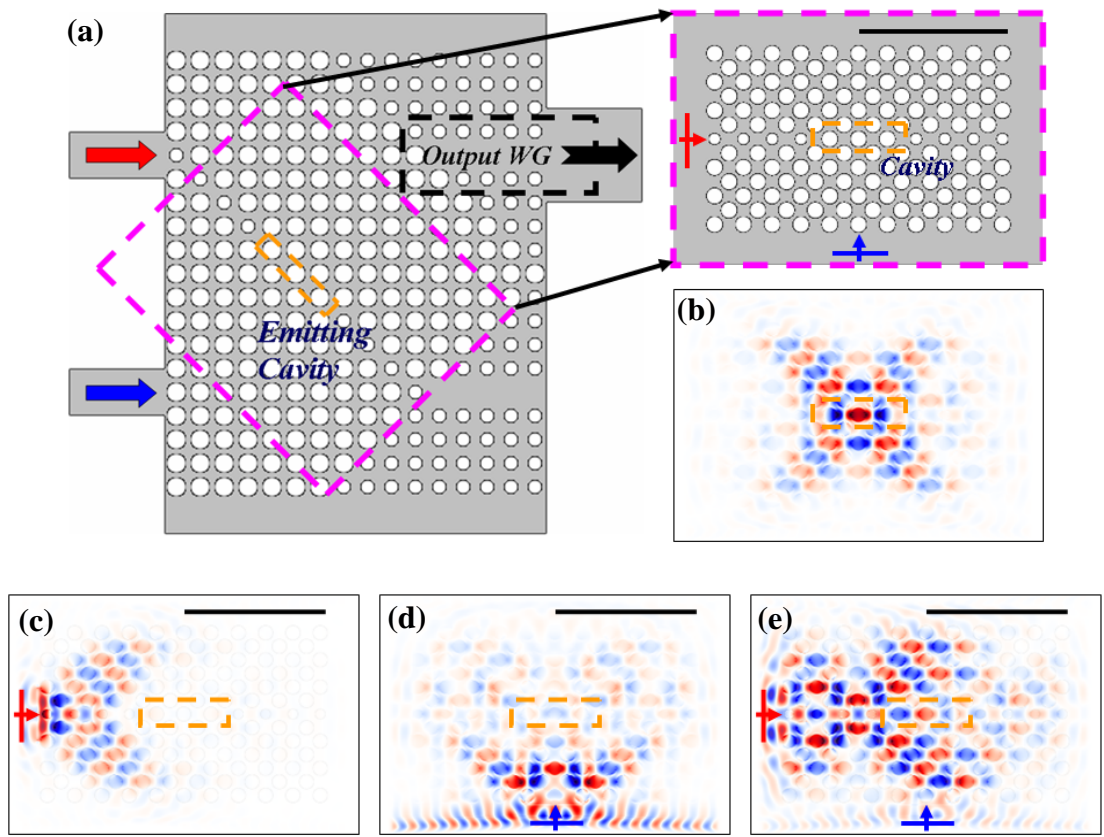

**Figure 3 | Principle of the optical AND logic gate.**
(a) Schematic geometry of the emitting cavity and the output waveguide components of the AND logic gate. (b) Calculated TE-like mode $E_y$ field distribution of the cavity resonance at the peak 0.2818 ($a/\lambda$). (c)-(e) Calculated $E_y$ field distribution of the light propagation at 0.2818 ($a/\lambda$) with single input [panels (c) and (d)] and with both two inputs [panel (e)], respectively. The emitting cavity that connects the two diodes chooses the leaky branch that enhances the output signal when two ports are both active and then outputs the AND signal from one of the diode waveguides.

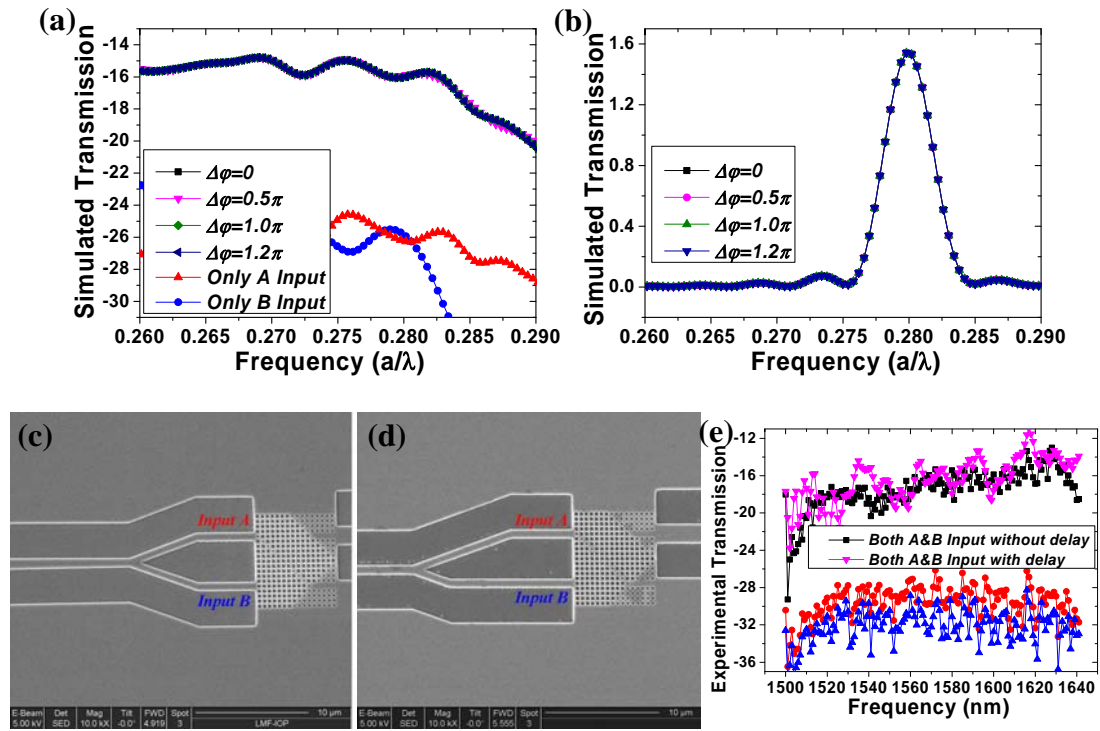

**Figure 4 | Phase properties of the AND logic gate.**
(a) Simulated transmission spectra of the AND logic with both two input Gaussian waves with different phase delays. (b) Simulated transmission spectra of the AND logic with both two input continuous waves with different phase delays. (c) Scanning electron microscope image of AND logic gate structure without phase delay. (d) Scanning electron microscope image of AND logic gate structure with arbitrary phase delay. (e) Experimental transmission spectra of AND logic gate structure with different phase delays. The experimental result fully supports the simulation results and shows that the logic gate has remarkable tolerance of phase sensitivity.

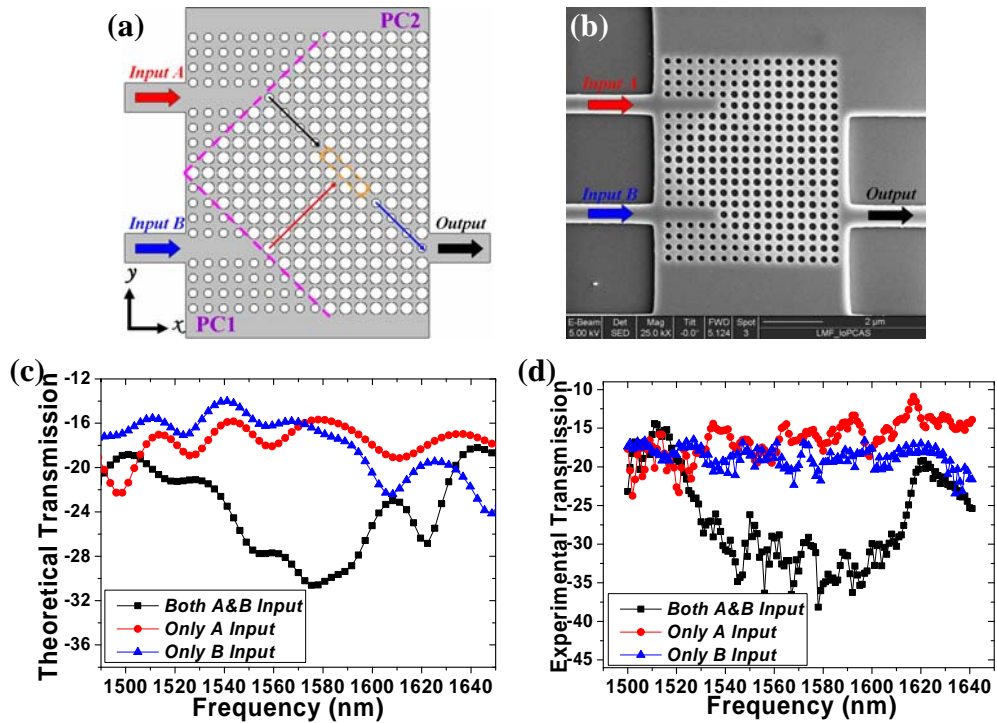

**Figure 5 | Demonstration of the optical NAND logic gate.**
(a) Schematic geometry and (b) scanning electron microscope images of the NAND logic gate. (c) Theoretical and (d) experimental transmission spectra of the NAND logic gate. The measured transmission spectra, showing apparent NAND logic transport of light, agree well with numerical simulations as the emitting cavity that connects two diodes here chooses different leaky branch that decreases the output signal when two ports are both active.